\documentclass[sigconf]{acmart}

\AtBeginDocument{%
  }

\setcopyright{acmlicensed}
\copyrightyear{2018}
\acmYear{2018}
\acmDOI{XXXXXXX.XXXXXXX}
\acmConference[Conference acronym 'XX]{Make sure to enter the correct
  conference title from your rights confirmation email}{June 03--05,
  2018}{Woodstock, NY}

\renewcommand\footnotetextcopyrightpermission[1]{}
\usepackage{multirow}
\usepackage{algorithm}
\usepackage{algorithmic}
\usepackage{booktabs}
\usepackage{xcolor}
\usepackage{amsmath}
\usepackage{utfsym}
\usepackage{epsfig}
\usepackage{graphicx}
\usepackage{dblfloatfix}
\usepackage{transparent}
\usepackage{listings}
\usepackage{graphicx}
\usepackage{subcaption}
\usepackage{makecell}

\begin{document}

\title{HLLM-Creator: Hierarchical LLM-based Personalized Creative Generation}

\author{Junyi Chen}
\authornote{Equal contribution. $^\text{\textdagger}$ Corresponding author.}
\email{chenjunyi.s@bytedance.com}
\affiliation{%
  \institution{ByteDance}
  \country{}
}

\author{Lu Chi}
\authornotemark[1]
\email{chilu@bytedance.com}
\affiliation{%
  \institution{ByteDance}
  \country{}
}

\author{Siliang Xu}
\authornotemark[1]
\email{xusiliang@bytedance.com}
\affiliation{%
  \institution{ByteDance}
  \country{}
}

\author{Shiwei Ran}
\email{ranshiwei@bytedance.com}
\affiliation{%
  \institution{ByteDance}
  \country{}
}

\author{Bingyue Peng}
\email{bingyue.peng@bytedance.com}
\affiliation{%
  \institution{ByteDance}
  \country{}
}

\author{Zehuan Yuan$^\text{\textdagger}$}
\email{yuanzehuan@bytedance.com}
\affiliation{%
  \institution{ByteDance}
  \country{}
}

\renewcommand{\shortauthors}{Trovato et al.}

\begin{abstract}
AI-generated content (AIGC) technologies are increasingly used for content creation across diverse domains. However, most current AIGC systems depend heavily on the inspiration of content creators, with limited exploration into generating truly personalized content tailored to individual users. In real-world applications such as online advertising, a single product may have multiple selling points, with different users focusing on different features. This underscores the significant value of personalized, user-centric creative generation.
Effective personalized content generation faces two main challenges: (1) accurately modeling user interests and integrating them into the content generation process while adhering to factual constraints, and (2) ensuring high efficiency and scalability to handle the massive user base in industrial scenarios. Additionally, the scarcity of personalized creative data in practice complicates model training, making data construction another key hurdle.
To tackle these issues, we propose HLLM-Creator, a hierarchical large language model framework designed to efficiently model user interests and generate personalized creative content. During inference, a combination of user clustering and a user-ad-matching-prediction based pruning strategy is employed to significantly enhance generation efficiency and reduce computational overhead, making the approach suitable for large-scale deployment. Moreover, we design a data construction pipeline based on chain-of-thought (CoT) reasoning, which generates high-quality, user-specific creative titles and ensures factual consistency despite limited personalized data. This pipeline serves as a critical foundation for the effectiveness of our model.
Extensive experiments on personalized advertising title generation for Douyin Search Ads show the effectiveness of HLLM-Creator. Online A/B test results demonstrate the practical value and scalability of our approach, with Adss increasing by \textbf{0.476\%}, paving the way for more effective and efficient personalized content generation in real-world industrial applications.
Codes for academic dataset are available at \textbf{\url{https://github.com/bytedance/HLLM}}.
\end{abstract}
\vspace{-1em}

\ccsdesc[500]{Information systems~Online advertising}
\ccsdesc[500]{Information systems~Recommender systems}

\keywords{Personalized Creative Generation, Online Advertising, Recomendation System, Large Language Model}

\settopmatter{printfolios=true}
\maketitle

\section{Introduction}
\begin{figure}
    \centering
    \includegraphics[width=0.97\linewidth]{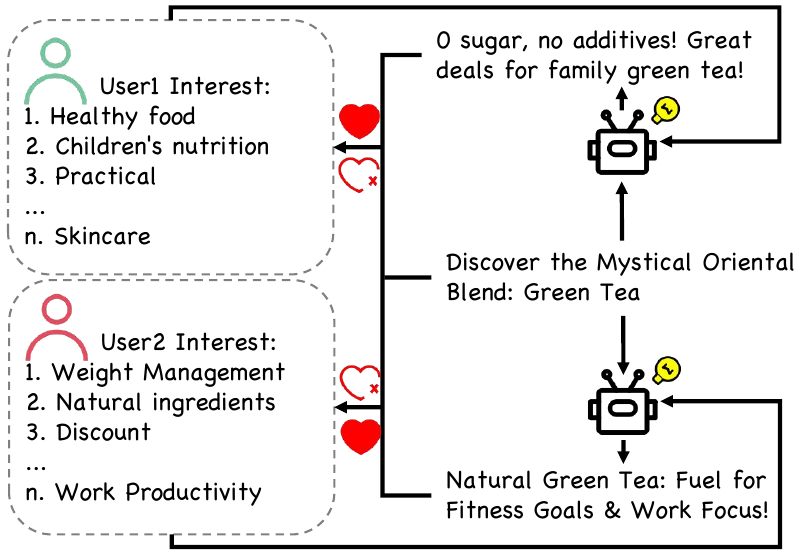}
    \caption{Personalized Creative Generation, which generates creatives more aligned with users' preferences, achieving "Different strokes for different folks".}
    \label{fig:top}
\end{figure}
The landscape of information dissemination and content creation has undergone dramatic changes over the past decades. In the early days of the internet, information was primarily published on static web pages by a limited number of administrators or companies, offering users little opportunity for interaction. The emergence and rapid evolution of recommender systems fundamentally shifted this paradigm: systems began proactively pushing content most likely to interest each user~\cite{rendle2010factorization,cheng2016wide,pi2020search}, while the barriers to content creation were significantly lowered. Platforms such as TikTok have enabled anyone to become a content creator, resulting in an explosion of user-generated content. More recently, advances in AI-generated content (AIGC) technologies~\cite{chatgpt,rombach2022high,liu2024sora,hurst2024gpt} have further democratized content creation, leading to a proliferation of AI-generated materials across the web. Despite these advancements, a critical gap remains: the precise fulfillment of individual users’ personalized needs is still far from being realized. Taking online advertising as an example, a single product often has multiple selling points, and different users may focus on different aspects. An example of a personalized advertising title is illustrated in Figure~\ref{fig:top}. For advertisers, it is extremely challenging to create finely tailored ads for diverse user groups—both due to the sheer scale of creative work required and the difficulty of accurately identifying user preferences in advance, especially for small and medium-sized advertisers. As a result, even when a product could meet a user’s needs, suboptimal ad descriptions may cause potential customers to be overlooked. Generating factually accurate, personalized ad descriptions is beneficial for all parties: advertisers can more precisely highlight product features and attract attention, users can quickly determine whether an ad matches their needs, and platforms can deliver content more accurately to relevant audiences. This underscores the importance and value of personalized creative generation.

Effective personalized creative generation must satisfy two key criteria: (1) accurately addressing user needs and pain points, and (2) strictly adhering to factual information, avoiding hallucinations and misleading “clickbait” content. However, most existing works on personalized generation have relatively weak user modeling capabilities, typically relying on simple user attributes (such as age and gender)~\cite{yang2024new} or keywords from users’ browsing history~\cite{cai2023generating,shen2024pmg} to represent user interests. 
Hallucination is unacceptable in the advertising domain. To tackle this challenge, previous works have constructed extensive factually grounded training datasets~\cite{mita2023striking}, ensured the completeness of input information to the model~\cite{wei2022creater,mita2023striking,lei2022plato}, and performed additional fine-tuning of generative models~\cite{wei2022creater,lei2022plato}. In our work, we largely follow this series of strategies and additionally deploy a strict hallucination detection mechanism in online settings to further mitigate risks.

Scalability and efficiency are also critical for real-world deployment. Industrial applications often serve hundreds of millions of users and manage millions of ads daily. Efficiently generating personalized ad titles for such a massive user base is a significant challenge. Existing approaches tend to either compromise on user modeling (sacrificing effectiveness)~\cite{yang2024new,shen2024pmg} or ignore the scale of the user base (sacrificing efficiency)~\cite{xu2025personalized,xu2025drc}.

To address these challenges, we present HLLM-Creator, a novel framework for personalized creative generation. Our approach leverages Hierarchical Large Language Models (HLLM~\cite{chen2024hllm}) to accurately extract user interests from historical behavior. HLLM consists of two LLMs: the Item LLM encodes ad titles into item representations, while the User LLM aggregates representations of ads previously clicked by the user to produce a user embedding. This hierarchical modeling strategy not only significantly improves training and inference efficiency compared to directly inputting raw click sequences into an LLM, but also fully exploits the world knowledge embedded in LLMs, achieving state-of-the-art performance in sequential recommendation.
Once the user embedding is obtained, it is combined with ad-side constraints (such as the original ad title and product selling points) and fed into a Creative LLM, which generates personalized ad titles in an autoregressive manner. The inclusion of user embeddings enables the Creative LLM to effectively perceive user interests, prioritize relevant selling points, and incorporate expressions likely to attract specific users (e.g., “a must-have for office workers”). At the same time, ad-side constraints guide the model to rewrite only based on the provided information, thereby ensuring factual consistency and minimizing hallucination.

Generating a unique personalized title for every ad-user pair is impractical in industrial settings. To maximize personalization benefits within resource constraints, we design two strategies. First, we cluster user embeddings into groups, with the number of clusters determined by available inference resources. This assumes that users within a cluster share similar interests, allowing a single generated title to serve many users while still meeting personalization needs. Second, since each ad is typically relevant to only a subset of user groups, we introduce an ad-user matching model to predict relevance scores between each ad and all user cluster centers. We then generate personalized titles only for the top-matched clusters, optimizing resource utilization and maximizing ROI.

Beyond modeling, training data quality is another critical factor. 
In practice, advertisers often focus on making their titles broadly appealing by stacking multiple attractive selling points, but they rarely create distinct titles for different user groups.

This lack of personalization makes original advertiser titles suboptimal for training personalized models. Furthermore, the high cost of manual annotation precludes large-scale labeled datasets. Consequently, leveraging synthetic data becomes a necessary alternative.

Some prior works attempt to leverage the generalization ability of generative models to produce personalized creatives by feeding user information into the model, but our experiments show that the quality of generated content is highly sensitive to prompt design. Meanwhile, synthetic data can lead to severe hallucination issues.
To address this, we carefully design prompts and employ a chain-of-thought (CoT) approach to construct high-quality, user-personalized training data, substantially raising the performance ceiling.
Furthermore, rigorous data cleaning procedures are applied to the constructed dataset to ensure the absence of hallucination issues in the training data.

Finally, our method has been deployed in the Douyin Search Ads Platform, a real-world industrial environment, generating personalized ad titles for hundreds of millions of users and delivering statistically significant improvements in key metrics through A/B testing. In summary, our main contributions are as follows:
\begin{itemize}
\item We propose a novel hierarchical LLM framework for personalized creative generation, enabling precise user interest modeling and fine-grained content personalization.
\item A Chain-of-Thought-based data construction pipeline is developed to expand personalization space and ensure factual consistency, effectively reducing hallucinations in generated titles.
\item A flexible and efficient inference scheme is developed for large-scale industrial deployment, with significant positive results in Douyin search advertising demonstrating its real-world impact.
\end{itemize}

\section{Related Work}

\begin{figure*}
  \centering
  \includegraphics[width=0.99\linewidth]{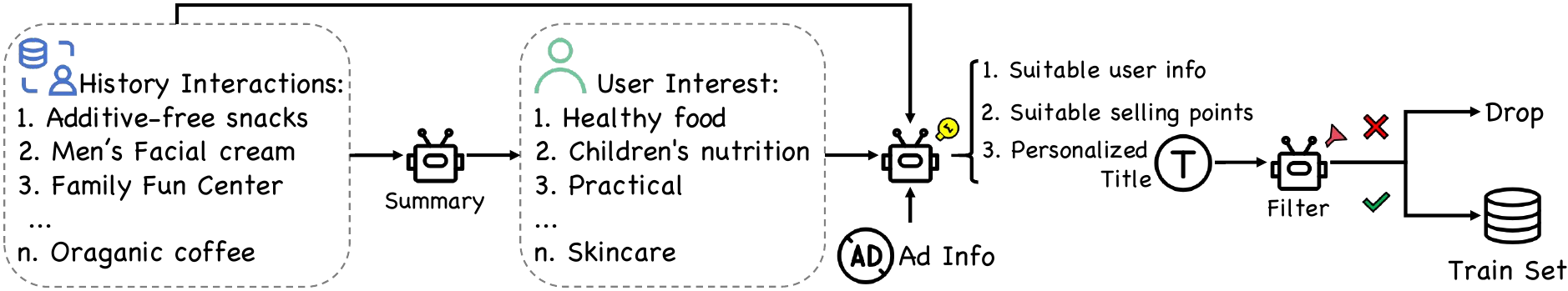}
  \caption{CoT-driven personalized title dataset construction pipeline. The LLM first summarizes user interests based on user historical sequences. Then it takes the user sequences, user interests, and advertising information as input to generate personalized titles. Finally, a strict LLM-based filtering mechanism is employed to remove hallucinated titles.}
  \label{fig:data_pipeline}
\end{figure*}

With the advancement of AIGC technology, it has been widely adopted in various domains, including text generation~\cite{gu2020generating,kanungo2021ad,mita2023striking,wei2022creater,vasudevan2025llm,lei2022plato,liu2025llms,chen2025carts,qiu2025measuring} and image generation~\cite{li2023planning,czapp2024dynamic,chen2024virtualmodel,gao2025postermaker}. These technological breakthroughs have revolutionized content production by significantly reducing material creation costs.
To align AIGC with user behavior objectives (e.g., click-through rate, CTR), several studies have attempted to integrate AIGC with reinforcement learning (RL) techniques. These efforts utilize real user behaviors as feedback signals to optimize generative models~\cite{wei2022creater} or train reward models based on user actions to guide the refinement of generative processes~\cite{kanungo2022cobart,zeng2023let,zhou2024gcof,chen2025ctr,lin2025sell}. While such methods have improved user click rates, they tend to generate high-performing texts or images tailored to the global user base, catering only to the preferences of the majority (i.e., the "head" users) rather than addressing the unique needs of individual users.

To address the limitations of previous methods, several personalized generation approaches have emerged~\cite{yang2024new,cai2023generating,meguellati2024good,tang2024step,xu2025drc,xu2025personalized,xu2024diffusion,shen2024pmg}. 
Despite their contributions, these personalized generation methods suffer from two key limitations: inadequate user modeling or inapplicability to large-scale industrial inference scenarios. For example, the personalized prompt model in CG4CTR~\cite{yang2024new} relies solely on simplistic user attributes (e.g., age, gender), which have been proven insufficient for modeling user interests in recommendation systems~\cite{zhou2018deep,pi2020search}. The approach in~\cite{cai2023generating}, which summarizes user behaviors into a few keywords, incurs significant information loss and heavily depends on the quality of keyword extraction. While PMG~\cite{shen2024pmg} addresses the limitations of keywords by introducing soft preference embeddings—an idea somewhat similar to ours—we observe that direct end-to-end training is challenging to optimize. To mitigate this, we incorporate multiple auxiliary losses and leverage HLLM~\cite{chen2024hllm} for more effective modeling of user behavioral features.
Regarding industrial deployment, existing works lack practical considerations. PMG~\cite{shen2024pmg}, for instance, is difficult to deploy in scenarios with hundreds of millions of users. Pigeon~\cite{xu2025personalized} and DRC~\cite{xu2025drc} adopt target-aware modeling approaches that result in a user-advertisement ($|\text{user}|\times|\text{ad}|$) scale, making them computationally prohibitive. In contrast, our work explicitly addresses industrial deployment requirements by designing clustering and pruning strategies, ensuring greater flexibility and efficiency in large-scale industrial settings.

\section{Method}
In this section, we first provide the problem formulation for the personalized creative generation task, followed by a description of the data construction pipeline (Section~\ref{sec:dataset}), which is one of the key factors for the effectiveness of our approach. Finally, we present the overall architecture of HLLM-Creator (Section~\ref{sec:architecture}), training objectives (Section~\ref{sec:training}), and inference strategies (Section~\ref{sec:serving}) for industrial deployment.

\subsection{Problem Formulation}

Given a user $u$ and target ad $I$, we aim to learn a generation model $p_{\Theta}(y|u, I)$ that can generate a personalized creative $y$ that aligns more closely with user interests (where $\Theta$ denotes trainable parameters of the model). More specifically, $y$ refers to the personalized advertising title, and $u$ is modeled with the user behavior sequence $H_u=\{I_1,I_2,...,I_n\}$, where $n$ denotes the sequence length. The available information of $I$ includes the original ad title $y_{orig}$ and the set of selling points $\mathbf{s}$.

\begin{figure*}
  \centering
  \includegraphics[width=0.99\linewidth]{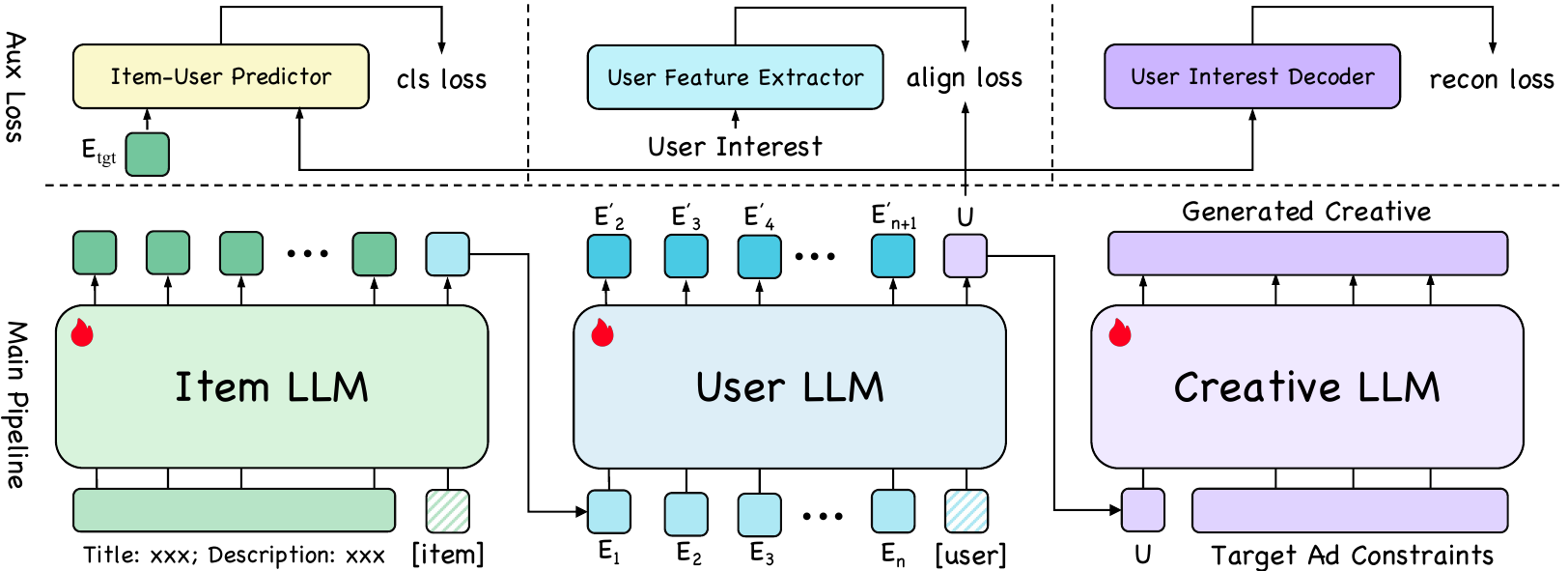}
  \caption{Overview of HLLM-Creator training framework. HLLM-Creator includes three LLMs: Item LLM, User LLM, and Creative LLM. Item LLM and User LLM are used to model user interests, while Creative LLM generates personalized titles for different users. Additionally, three auxiliary losses are added to enhance the extraction of user interests.}
  \label{fig:train}
\end{figure*}

\subsection{CoT-driven Personalized Title Dataset Construction}
\label{sec:dataset}
In real-world scenarios, advertisers rarely have the capacity to create personalized titles for distinct user groups, leading to a scarcity of personalized data for model training. To address this challenge, we leverage LLM (e.g., DeepSeek-R1~\cite{guo2025deepseek}) to construct synthetic personalized datasets for training purposes. A naive approach would involve feeding user behavior sequences, the target ad title, and key selling points into the LLM and prompting it to directly generate personalized titles. However, we found that titles generated through this method exhibit limited personalization (experimental results can be found in Section~\ref{sec:exp_data}). Inspired by chain-of-thought (CoT)~\cite{wei2022chain} techniques widely adopted in the field of LLM, we decompose the personalized title generation process into several key subtasks and leverage the intermediate results of these subtasks to ultimately generate high-quality personalized titles. Figure~\ref{fig:data_pipeline} illustrates the proposed CoT-driven high-quality dataset construction pipeline. All prompts are provided in the Appendix~\ref{sec:prompt}.

\subsubsection{User Interest Profiling}
We first instruct the LLM to extract predefined multi-dimensional user interests (such as long-term interests, short-term preferences, and specific needs) from user behavior sequences \( H \). Then, we prompt the LLM to generate personalized titles based on these user profiles, combined with the original ad title and ad selling points.

\subsubsection{Interest-Driven Title Generation}
Regarding the forms of user personalization in title generation, we hypothesize they can be categorized into two types: one is directly incorporating descriptions targeting specific user groups (e.g., "a must-have for office workers"), and the other is integrating advertising selling points that users may be more interested in.
Therefore, during the generation of personalized titles, we guide the LLM to first extract user information suitable for integration into the current advertisement based on the provided user interests, then identify the selling points that the user is more likely to care about, and finally these elements are combined to generate a personalized ad title. 
This CoT-based generation method reduces the processing difficulty for the LLM, allows for more thorough consideration of user interests and advertising selling points, and thus generates higher-quality titles.
\subsubsection{Hallucination-Free Title Filtering}
Synthetic data inevitably suffers from hallucination issues (i.e., fabricating non-existent information), even if we emphasize that the LLM should avoid fabricating information out of thin air during the synthesis process. Although titles with hallucinations may offer superior personalization, such "clickbait-style titles" are strictly prohibited in industrial scenarios. Therefore, we employ the LLM to implement a strict filtering mechanism on the generated personalized titles. It can effectively mitigate this critical issue by ensuring strict adherence to the source material.

\begin{figure*}
  \centering
    \includegraphics[width=0.99\linewidth]{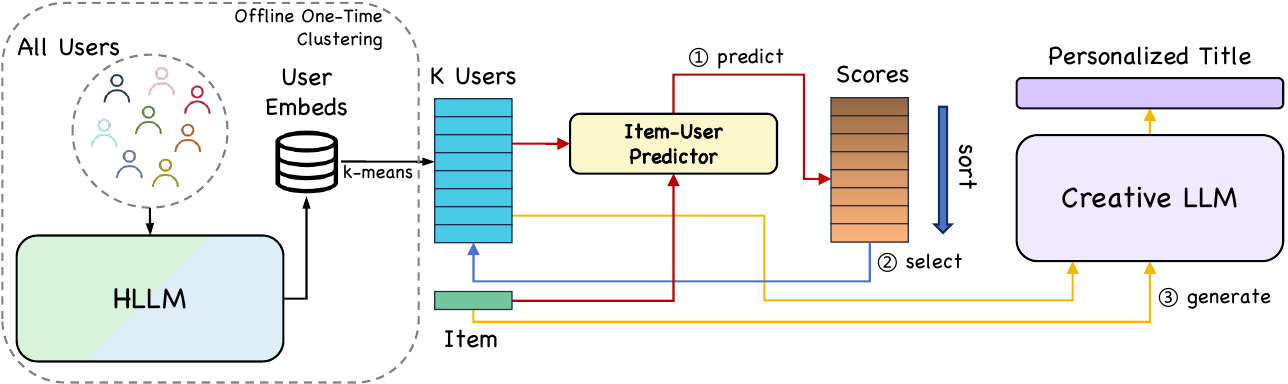}
    \caption{Infrence workflow of HLLM-Creator. All users first have their user embeddings extracted via HLLM, followed by offline one-time clustering into $K$ clusters. During actual inference, the top-$k$ user clusters that best match the ad are selected to generate personalized titles.}
    \label{fig:serving}
\end{figure*}

\subsection{Model Architecture}
\label{sec:architecture}
The model architecture and training process of HLLM-Creator are shown in Figure~\ref{fig:train}.
HLLM-Creator consists of three LLMs in total: Item LLM is used to extract item features, which serve as input to the User LLM; User LLM adopts a late fusion structure as introduced in HLLM~\cite{chen2024hllm} for extracting user features; and Creative LLM is used for generating personalized content.

Specifically, for each \( I \in H \), we first flatten its corresponding textual attributes into the sentence $\mathcal{T}$.
After passing through the LLM tokenizer, we append a special token \texttt{[item]} at the end, thus the input token sequence for the Item LLM can be formulated as $\{t_1, t_2, \dots, t_m, \texttt{[item]}\}$ where $m$ represents the length of item text tokens. The hidden state from the last layer corresponding to the special token \texttt{[item]} is considered as the item embedding $E$.

Then the original user history sequence $H = \{I_1, I_2, \dots, I_n\}$ can be transformed into a historical feature sequence $E = \{E_1, E_2, \dots, E_n\}$ through the Item LLM, where $E_i$ represents the item embedding of $I_i$. 
Similar to the Item LLM, we append a special \texttt{[user]} token after the feature sequence \( E \) as input to the User LLM, and take the hidden state corresponding to the \texttt{[user]} token as the user embedding $U$.

The Creative LLM first converts the original text information of advertisements (such as title, selling points, etc.) into corresponding embeddings \( F = \{{ f_1, f_2, \dots, f_a \}} \) through the word embedding layer, where $a$ represents the length of the ad text tokens. These embeddings are then concatenated with the user embedding \( U \) to form the input sequence \( \{{ U, f_1, f_2, \dots, f_a \}} \) for the Creative LLM, enabling the model to generate personalized titles tailored to the user’s interests.

\subsection{Training Objectives}
\label{sec:training}
\subsubsection{Generative Loss}
The main training objective of HLLM-Creator is consistent with conventional Supervised Fine-tuning (SFT). It supervises the output of Creative LLM to align with the synthetic high-quality personalized titles introduced in Section~\ref{sec:dataset}, adopting the paradigm of next token prediction for training.
The generative loss function can be formulated as:
\begin{equation}
  \mathcal{L}_{gen} = -\frac{1}{L} \sum_{i=1}^{L} \log\left(p\left(r_j \mid U, f_1, f_2, \dots, f_a, r_1, \dots, r_{i-1} \right)\right)
\end{equation}
where $r$ (response) denotes the synthesized training data, and $L$ is the length of the response text tokens.

However, using only the training objective of personalized title generation for end-to-end training of all model parameters yields only suboptimal results. We attribute this to the fact that optimizing the User LLM for extracting user embeddings is relatively challenging. Therefore, three auxiliary losses are introduced in the following sections to explicitly supervise the user embedding extraction.

\subsubsection{Recommendation Objective Loss (cls loss)}
A user feature that adequately models user interests should be sufficiently capable of judging user preferences. 
Therefore, referring to the late fusion version of HLLM, we introduce a classification loss for whether the user clicks on the candidate item to supervise the training of user features. 
Here, positive examples are items that were actually clicked by the user, while negative examples are randomly sampled. 
Specifically, we input the user feature $U$ and the candidate item feature $E$, which is extracted by the Item LLM, into the Item-User Predictor, implemented by a shallow fully connected neural network, to output the click probability.
The cls loss function can be formulated as:
\begin{equation}
\begin{aligned}
\mathcal{L}_{\text{cls}} &= -\frac{1}{M} \sum_{i=1}^{M} \left[ y_i \log(\sigma(f_{\text{dense}}(U_i, E_i))) \right. \\
&\quad + \left. (1-y_i) \log(1 - \sigma(f_{\text{dense}}(U_i, E_i))) \right]
\end{aligned}
\end{equation}
where \(f_{\text{dense}}\) denotes the Item-User Predictor, \(U_i\) and \(E_i\) represent user embedding and item embedding respectively, \(\sigma\) is the sigmoid function. $y$ indicates whether the user clicked, and $M$ is the total number of positive and negative samples in a training batch.

\subsubsection{Semantic Alignment Loss (align loss)}
Here, we use the user interests extracted in Section~\ref{sec:dataset} to assist in supervising the learning of $U$.
We use an LLM as the User Feature Extractor, and append a special token after the user interests text description to extract interest features $V$. Contrastive learning is performed between $V$ and the user embeddings $U$ generated by the User LLM, using the InfoNCE~\cite{oord2018representation} loss to pull closer the two types of features from the same user and push apart the two types of features from different users.
The align loss function can be formulated as:
\begin{equation}
\mathcal{L}_{\text{align}} = -\frac{1}{N} \sum_{i=1}^{N} \log \frac{\exp(\text{sim}(U_i, V_i)/\tau)}{\sum_{j=1}^{N} \exp(\text{sim}(U_i, V_j)/\tau)}
\end{equation}
where $\text{sim}$ denotes a cosine similarity function, \(\tau\) is temperature, and $N$ is the number of samples in a training batch.

\subsubsection{Reconstruction Loss (recon loss)}
We hypothesize that user embeddings should contain compressed user interest information. To achieve this, a decoder (LLM) is connected after the user embeddings, where the supervision target for the decoder output is the user interest description extracted in Section~\ref{sec:dataset}, and the training follows the next token prediction paradigm.
The recon loss function can be formulated as:
\begin{equation}
\mathcal{L}_{\text{recon}} = -\frac{1}{T} \sum_{t=1}^{T} \log P(w_t | U, w_1, \ldots, w_{t-1})
\end{equation}
where $w$ denotes the user interest description, and $T$ is its text length.

The overall training objective of HLLM-Creator is:
\begin{equation}
\mathcal{L} = \mathcal{L}_{\text{gen}} + \lambda_{\text{cls}} \mathcal{L}_{\text{cls}} + \lambda_{\text{align}} \mathcal{L}_{\text{align}} + \lambda_{\text{recon}} \mathcal{L}_{\text{recon}}
\end{equation}
where each \(\lambda\) represents the weight of the corresponding loss.

\subsection{Inference workflow}
\label{sec:serving}
Generating personalized titles for every ad-user pair is impractical in industrial scenarios. Two strategies are designed during inference to maximize the benefits of personalized creative generation under limited resources.

As shown in Figure~\ref{fig:serving}, we first use HLLM (Item LLM + User LLM) to extract a large number of user embeddings, and then perform k-means clustering based on these user embeddings to categorize users into $K$ groups.
During inference, the cluster center embedding $U_k$ is fed into the Creative LLM as the user embedding, while other ad information remains unchanged. While clustering may compromise some granularity of personalization, it enables a single generation to cover more users, thereby enhancing production efficiency. Related experiments can be found in Section~\ref{sec:exp_clustering}.

Furthermore, usually each ad has limited target audience, most ads will not reach all user groups. 
Therefore, newly created ads do not need to generate personalized titles for all $K$ clusters. 
Instead, we introduce a predictor to determine the matching score between user groups and advertisements, and only perform personalized generation for user groups with a high matching score. 
Here we reuse the Item-User Predictor shown in Figure~\ref{fig:train}. 
In the inference phase, the $K$ cluster center embeddings are fed into the Item-User Predictor with the current ad embedding $E_{tgt}$ respectively to determine their matching degree with the current ad. 
Based on these scores, the top-$k$ user clusters are selected for personalized title generation.

\clearpage

\section{Experiments}
HLLM-Creator is validated on Douyin Search Ads Platform.

\subsection{Implementation Details and Evaluation Setup}
\subsubsection{Dataset Construction}
An industrial dataset from Douyin Search Ads is adopted for training and evaluation. 
First, we selected 2.6 million samples from online user click logs, with each sample consisting of the user's click sequence, search query, original ad title, and selling points. Then, personalized titles were generated using the CoT-driven data construction method introduced in Section~\ref{sec:dataset}. After filtering out personalized titles with hallucinations, approximately one quarter of the data remained.
The filtered data are split into 650K samples for model training and 500 samples for offline evaluation.

\subsubsection{Training Configuration}
TinyLlama-1.1B~\cite{tinyllama} is adopted for both the Item LLM and the User LLM, while the Creative LLM employs Qwen3-8B~\cite{yang2025qwen3}.
The user's historical click sequence is truncated to a maximum length of 500. Only the title with a maximum of 64 tokens is used as input to the Item LLM.
All auxiliary loss weights are set to 1. 
The user embedding $U$ is aligned to the dimension of the Creative LLM through a single-layer fully connected network. 
The Item-User Predictor is an 8-layer MLP-Mixer~\cite{tolstikhin2021mlp} with a hidden dimension of 2048.
To save GPU memory, the parameters of the User Feature Extractor, User Interest Decoder, and Creative LLM are shared.
All parameters are initialized with the pre-trained LLM weights and trained end-to-end for one epoch with a learning rate of 2e-5.
Unless otherwise specified, all experimental results for HLLM-Creator are based on clustering into 256 groups.

\subsubsection{Incorporating Search Query}
Our evaluation scenario is primarily the search scenario, where the user's search query is important information that represents the user's immediate interests. Therefore, reflecting the user's search query in the title is very important. 
Our method is highly flexible in incorporating new constraints, with two main upgrades: First, we introduce the user's search query during data construction (the specific prompt can be found in the Appendix~\ref{sec:prompt}); second, we update the input to the Creative LLM by appending the search query after the user embedding $U$ and ad information.

\subsubsection{Reproducibility}
To facilitate the reproduction and validation of the effectiveness of HLLM-Creator, we conducted experiments on the academic dataset (Amazon Book Reviews~\cite{books}). The experimental results are provided in the Appendix~\ref{sec:academic}, and the relevant codes are available at \textbf{\url{https://github.com/bytedance/HLLM}}. 

\subsubsection{Evaluation Metrics}
\label{sec:eval_metric}
\paragraph{Online Evaluation Metrics}
We use three key metrics: Adss (Advertiser Score), Advv (Advertiser Value), and RankAdvv (Rank Advertiser Value) for online evaluation. The latter two are defined as:
\begin{align}
\text{Advv} &= \text{cpa\_bid} \times \text{conversions} \\
\text{RankAdvv} &= \text{rank\_bid} \times \text{conversions}
\end{align}
where \text{cpa\_bid} denotes the cost-per-action bid, and \text{rank\_bid} is the bid used by the ads ranking system.

\paragraph{Offline Evaluation Metrics}
Directly quantifying the degree of user interest in different titles is inherently challenging.
We adopt the Good-Same-Bad (GSB) preferences of GPT4.1~\cite{openai2025gpt41}. Pormpt can be found in Appendix~\ref{sec:prompt}. 
To avoid order bias in LLM evaluation, we test both (A, B) and (B, A) prompt orders. Only consistent results are counted; otherwise, the outcome is marked as "Same". We define "advantage" as follows:
\begin{equation}
\text{Advantage} = \frac{N_{\text{Good}} - N_{\text{Bad}}}{N_{\text{Good}}+N_{\text{Same}}+N_{\text{Bad}}}
\end{equation}

For hallucination, we set up a hallucination detection pass rate metric. Using the same prompt as in the data cleaning process, we perform hallucination detection on the generated titles and calculate the proportion of titles without hallucinations. A higher value for this metric indicates a stronger ability of the model to adhere to factual information.
\subsection{Online A/B Test}

\begin{figure}
    \centering
    \includegraphics[width=0.98\linewidth]{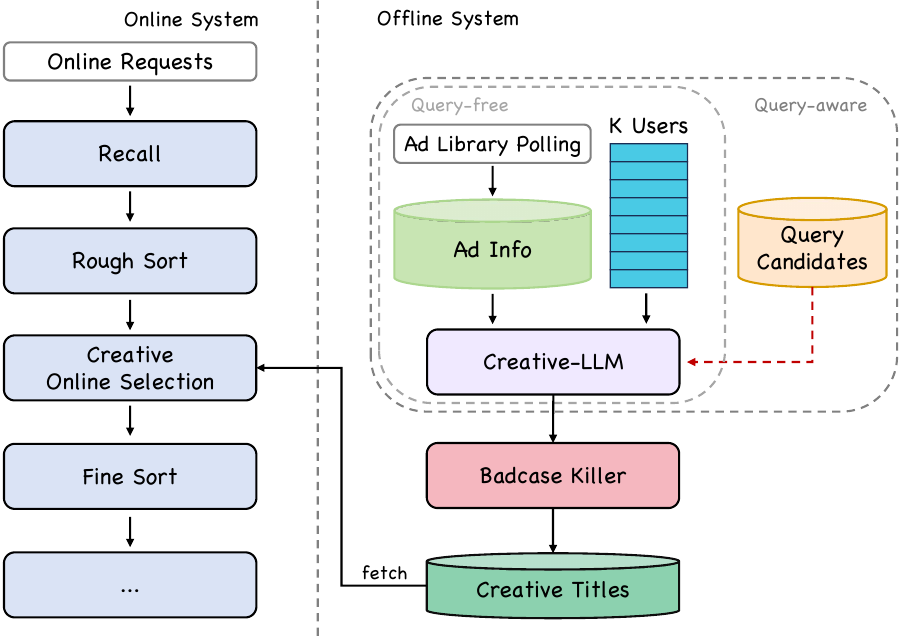}
    \caption{Industry deployment framework. See the main text for details.}
    \label{fig:enter-label}
\end{figure}

HLLM-Creator is validated on Douyin Search Ads Platform, a real-world industrial recommendation system. As shown in Figure~\ref{fig:enter-label}, our online A/B testing deployment consists of two key components: offline generation and online selection.

\subsubsection{Offline Generation}

Due to the inherent latency of LLM inference, real-time generation of personalized content in an online environment is impractical. To address this limitation, we adopt an offline inference strategy, where personalized titles are pre-generated in batches as described in Section~\ref{sec:serving}. Specifically, we cluster user embeddings at the million scale into 256 groups, and then generate personalized titles for each ad campaign. Two generation strategies are adopted for each ad campaign: (1). \textbf{Query-Aware}: Personalized titles are generated by incorporating the ad’s potential queries (based on historical statistics or model predictions). These titles fully leverage the relevance among the user, query, and ad. Considering resource constraints, this strategy generates titles only for the top-1 user group predicted by the Item-User Predictor. (2). \textbf{Query-Free}: Personalized titles are generated using only user and ad information, targeting the top-5 user groups. This approach enables coverage of a broader range of user groups.

We continuously poll and generate content for all active advertising campaigns across the entire library, with each update cycle taking around 5 hours. Generated titles are then processed by a rule-based filter, referred to as the Badcase Killer, which removes undesirable cases such as fabricated content, inappropriate numbers, locations, brands, and sensitive terms. The qualified titles will be stored in the creative library and used as candidates for material selection during online serving.

\subsubsection{Online Selection}
The online request pipeline remains largely unchanged. For each search request, relevant ad campaigns are recalled and go through a coarse ranking stage. After that, they enter the creative online selection module, where title selection is performed for each campaign. The personalized titles are added as new candidates to participate in the selection process. The winning title is then used as the title for the corresponding ad campaign in subsequent fine-ranking and other downstream stages.

\subsubsection{A/B Test Result}

\begin{table}
  \caption{Douyin Search Ads A/B Test Results.}
  \label{tab:ab_test_label}
  \centering
  \begin{tabular}{llll}
    \toprule
    Adss$\uparrow$ & Advv$\uparrow$ & RankAdvv$\uparrow$ \\ 
    \midrule
    \textbf{+0.476\%} & \textbf{+0.297\%} & \textbf{+0.51\%} \\
    \bottomrule
  \end{tabular}
  \begin{center}
  \end{center}
\end{table}

A/B test results can be found in Table~\ref{tab:ab_test_label}. Compared to the baseline, introducing personalized titles led to improvements of \textbf{0.476\%}, \textbf{0.297\%}, and \textbf{0.510\%} on the Adss, Advv, and RankAdvv metrics, respectively, which are considered good gains for a highly optimized system.  
It is worth noting that the baseline already includes various AI-Generated titles, including query-aware titles, which further demonstrate the benefits of introducing personalization. 

Additionally, we conducted an in-depth study of the Click-Through Rate (CTR) for personalized titles. During the A/B testing process, we marked ads where personalized titles won in the selection stage. For the corresponding baseline group, personalized titles were skipped, and the top-1 title was selected from the remaining titles for subsequent stages. By directly comparing the CTR performance of these ads between the baseline and experimental groups, the results showed that the experimental group achieved a \textbf{1.789\%} increase in CTR.

\subsection{Comparison with other Methods}

\begin{figure}
  \centering
  \includegraphics[width=\linewidth]{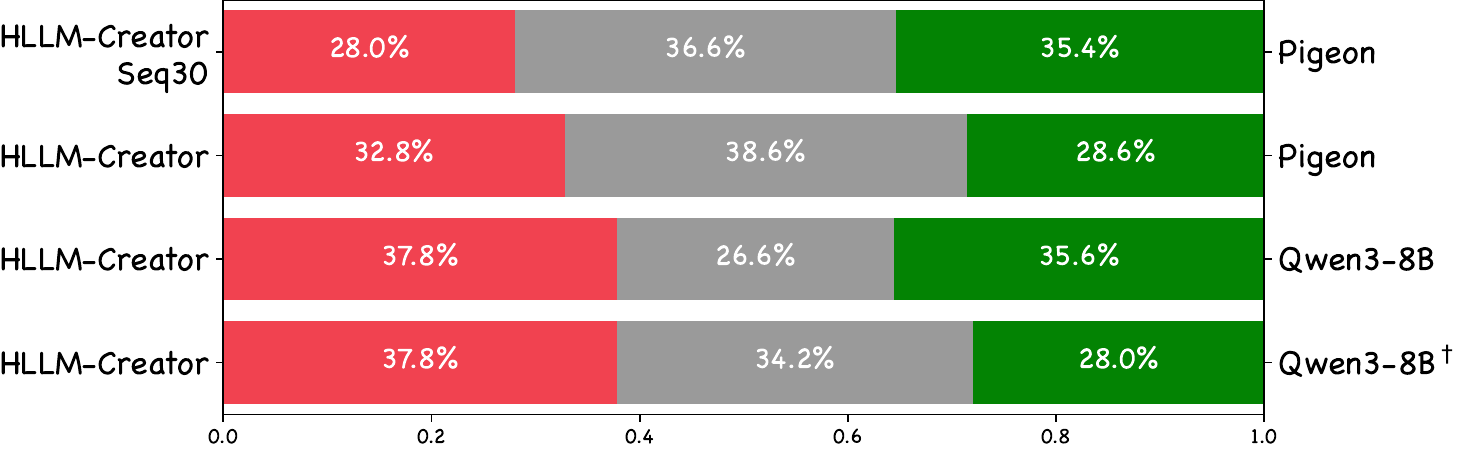}
  \caption{Comparison of HLLM-Creator with other baselines on LLM GSB evaluation. $\dagger$ indicates testing on the subset of titles from both methods that are free of hallucinations.}
  \label{fig:comparison}
\end{figure}

Figure~\ref{fig:comparison} shows the GSB results of HLLM-Creator compared with other methods.
Pigeon~\cite{xu2025personalized} is a recent state-of-the-art work on personalized generation. We reproduced this method on search ads data.
Pigeon is much less efficient than HLLM-Creator, mainly because it flattens the historical title sequences into a single sequence and its reference-aware modeling of historical behavior results in an inference complexity of $|\text{user}|\times|\text{ad}|$, whereas ours is only $K\times|\text{ad}|$. In terms of effectiveness, with the same sequence length, Pigeon has a 7.4\% advantage. However, when aligning training complexity, HLLM-Creator can extend the sequence length to 500, at which point it achieves a 4.2\% advantage.

We also compared our method with a general-purpose LLM trained on broad knowledge, by directly inputting user behavior sequences and user interests in text form into the LLM to generate personalized titles. For a fair comparison, we used Qwen3-8B~\cite{yang2025qwen3}—the same base model as our Creative LLM. The results show that our method has a slight 2\% advantage. Additionally, we observed that the LLM without finetuning performs poorly in terms of hallucination: as shown in Table~\ref{tab:Hallucination}, the hallucination detection pass rate for Qwen3-8B is 60\%, while our method achieves 75\%. When testing on the subset of titles from both methods that are free of hallucinations, the advantage of HLLM-Creator increases to 9.8\%.

\subsection{Ablation Study}

\subsubsection{Necessity of Personalization Modeling}
\begin{figure}
  \centering
  \includegraphics[width=\linewidth]{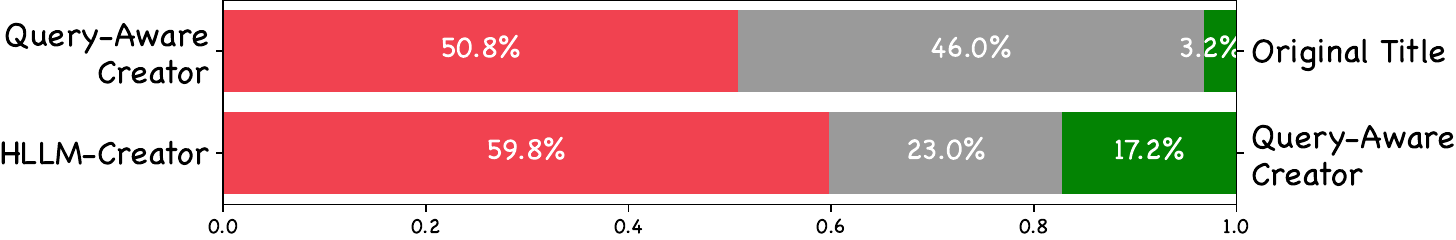}
  \caption{Ablation study on Personalization Modeling.}
  \label{fig:personal}
\end{figure}

We use the model trained on titles that are generated based solely on original ad information and user queries as the baseline, referred to as Query-Aware-Creator, to verify the necessity of further incorporating user modeling. As shown in Figure~\ref{fig:personal}, the Query-Aware Creator already achieves a significant improvement over the original titles. However, our HLLM-Creator is able to achieve further gains on top of this, demonstrating the necessity of personalization modeling.

\subsubsection{Effect of the CoT Process in Data Construction}
\label{sec:exp_data}
\begin{table}
  \caption{Ablation study on CoT-based personalized title generation in data construction pipeline.} 
  \label{tab:cot}
  \begin{tabular}{cccrrrr}
    \toprule
    \multirow{2}{*}{CoT 1}  & \multirow{2}{*}{CoT 2}       & \multirow{2}{*}{\makecell[c]{Chat\\Rounds}} & \multirow{2}{*}{Good} & \multirow{2}{*}{Same} & \multirow{2}{*}{Bad} & \multirow{2}{*}{Advantage$\uparrow$}  \\ 
    \\ \midrule
                & \usym{2713} & 1 & 33.0\% & 41.4\% & 25.6\% & 7.4\% \\
    \usym{2713} & \usym{2713} & 1 & 35.8\% & 42.0\% & 22.2\% & 13.6\% \\ 
    \usym{2713} &             & 2 & 45.0\% & 39.2\% & 15.8\% & 29.2\% \\
    \usym{2713} & \usym{2713} & 2 & 57.8\% & 28.0\% & 14.2\% & \textbf{43.6\%} \\ \bottomrule
\end{tabular}
\end{table}
There are two critical CoT steps in our data construction process: CoT1: User Interest Profiling; CoT2: Interest-Driven Title Generation. Ablation studies are shown in Table~\ref{tab:cot}. The baseline here does not involve any CoT process; instead, user behavior sequences, ad information, and query are directly input into the LLM to generate personalized titles in a single chat round. As shown, each CoT step brings significant performance improvements. Meanwhile, we attempted to combine the two CoT processes into a single dialogue, that is, requiring the LLM to first extract user interests, then identify key interests and selling points based on the ad information and user interests, and finally generate the title, all within one prompt. We found that, compared to the two-step dialogue, the single-step approach resulted in inferior performance.

\subsubsection{Hallucination Issue}
To address the hallucination problem, we implemented measures at both the data and model levels. On the data side, we used LLMs to perform hallucination detection on the constructed data and filtered out titles with hallucinations. On the model side, we provided sufficient original ad constraints in the input, including the original title and ad selling points. As shown in Table~\ref{tab:Hallucination}, each step contributes to alleviating the hallucination issue. Ultimately, we achieved a hallucination pass rate of 75\%.

\begin{table}
  \caption{Hallucination detection pass rates of different models.}
  \label{tab:Hallucination}
  \small
  \begin{tabular}{lcccr}
    \toprule
    \multirow{2}{*}{Method}  & \multirow{2}{*}{\makecell[c]{Data\\Cleaning}} & \multirow{2}{*}{\makecell[c]{Selling\\Points}} & \multirow{2}{*}{Titles} & \multirow{2}{*}{Pass$\uparrow$} \\ 
    \\ \midrule
    Qwen3-8B~\cite{yang2025qwen3} & - & & & 60\% \\
    HLLM-Creator & \usym{2717} & &  & 29\%  \\ 
    HLLM-Creator & & \usym{2717} & & 53\%  \\ 
    HLLM-Creator & & & \usym{2717} & 9\%  \\ 
    HLLM-Creator & & & & \textbf{75}\%  \\ \bottomrule
\end{tabular}
\end{table}

\subsubsection{Auxiliary Loss}

\begin{table}
  \caption{Ablation study on auxiliary losses.} 
  \label{tab:aux_loss1}
  \small
  \begin{tabular}{cccrrrr}
    \toprule
    align & cls & recon  & Good & Same   & Bad & Advantage$\uparrow$  \\ \midrule
                &             &             & 52.6\% & 25.6\% & 21.8\% & 30.8\% \\ 
    \usym{2713} &             &             & 56.8\% & 25.4\% & 17.8\% & 39.0\% \\ 
                & \usym{2713} &             & 56.4\% & 24.6\% & 19.0\% & 37.4\% \\ 
                &             & \usym{2713} & 53.8\% & 25.4\% & 20.8\% & 33.0\% \\ 
    \usym{2713} & \usym{2713} &             & 57.6\% & 24.4\% & 18.0\% & 39.6\% \\
    \usym{2713} & \usym{2713} & \usym{2713} & 59.8\% & 23.0\% & 17.2\% & \textbf{42.6\%} \\ \bottomrule
\end{tabular}
\end{table}

The impact of different auxiliary losses is explored in Table~\ref{tab:aux_loss1}. Here, the baseline is Query-Aware-Creator.
When trained without any auxiliary losses, the advantage of HLLM-Creator is only 30.8\%. Experiments show that regardless of which auxiliary loss is used, it is beneficial for generating personalized titles, with the advantage increasing by 2.2\% (30.8\% $\to$ 33.0\%) to 8.2\% (30.8\% $\to$ 39.0\%). Moreover, the effects of different auxiliary losses are complementary: when three aux losses are used simultaneously, the advantage of HLLM-Creator reaches  \textbf{42.6\%}.

\subsubsection{Clustering}
\label{sec:exp_clustering}
\begin{figure}
  \centering
  \includegraphics[width=0.93\linewidth]{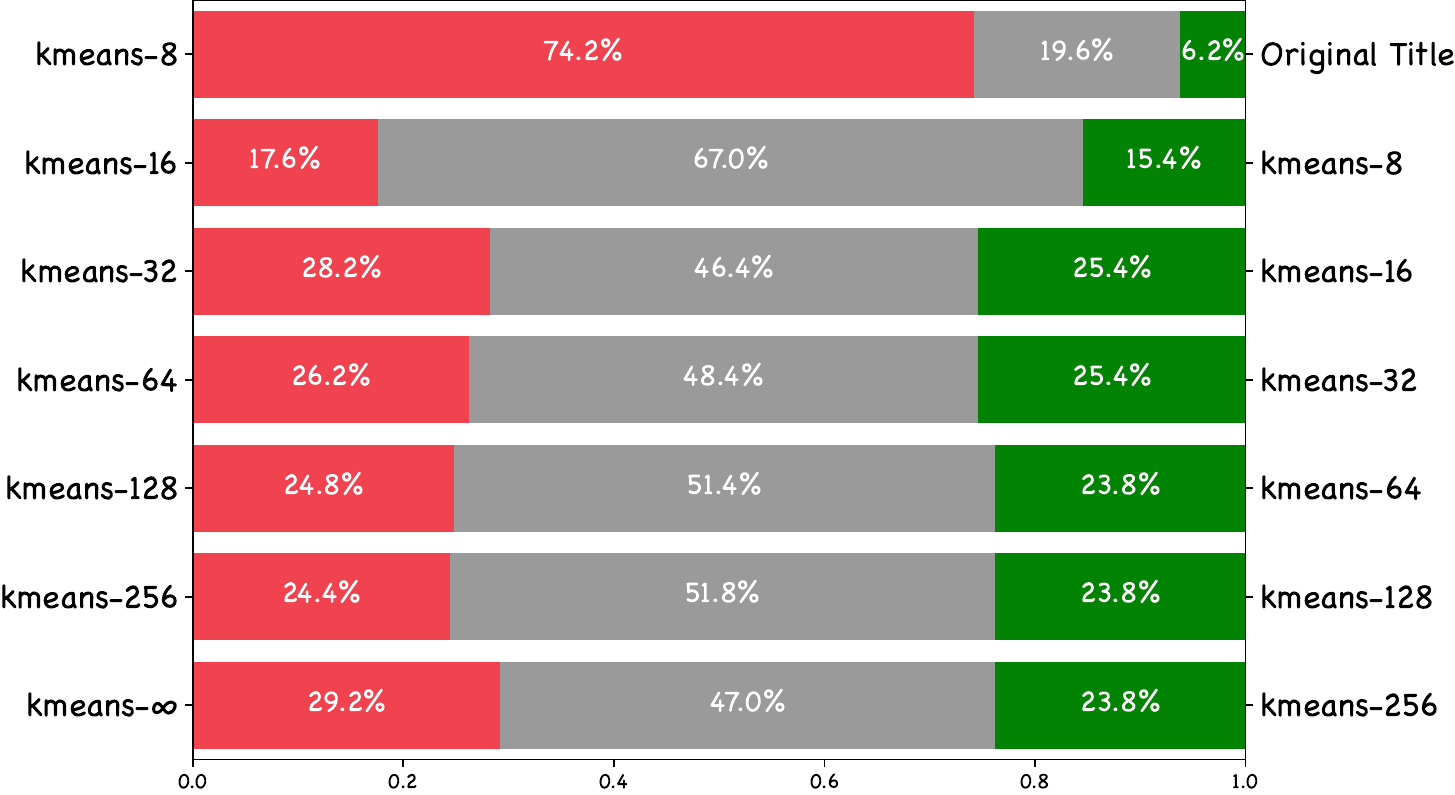}
  \caption{Ablation study on the impact of clustering. $\infty$ indicates no clustering applied.}
  \label{fig:clustering}
\end{figure}

To deploy HLLM-Creator in real-world industrial scenarios, we cluster all users before deployment to reduce inference costs. Clustering inevitably incurs clustering loss and affects the quality of generated personalized titles. Thus, we explore the impact of clustering on the performance in Figure~\ref{fig:clustering}.
It can be observed from Figure~\ref{fig:clustering} that even when clustered into a very small number of clusters (8), there is still a 68\% advantage compared to original titles. 
It can also be observed that as the number of clusters increases, the improvement in advantage becomes less significant. However, there is still considerable room for improvement compared to not using clustering. For example, compared to kmeans-256, the no-clustering setting kmeans-$\infty$ still achieves a 5.4\% advantage. This indicates that our method has a higher potential upper bound, and in the future work, we will explore the online ROI of scaling up the number of clusters.

\subsection{Case Study}
By examining the cases generated by HLLM-Creator, we found that they are generally consistent with our previous assumptions: the personalization of the generated titles is mainly reflected in two aspects—explicitly adding descriptions related to user groups and incorporating personalized selling points. 
For example, the generated titles may explicitly include descriptions such as "professional women" to highlight user characteristics. Alternatively, for users who are more concerned about quality of life, the model selects selling points like "health" and "smart," while discarding less appealing features such as "cost-effectiveness" or "large capacity" that may not attract the current user as much.

To protect the privacy of users and advertisers, we used LLMs to construct some illustrative data in Appendix~\ref{sec:case_study}, which is generally consistent with the experimental conclusions.

\section{Conclusion}
In this paper, we propose HLLM-Creator, an innovative personalized creative generation model based on hierarchical large language models. The model takes user historical behavior and ad-side constraint information as input and outputs personalized creatives. To enable deployment in real-world industrial environments, we designed clustering and pruning strategies based on the matching degree between ads and user interests. To address the lack of personalized data in real scenarios, we developed a CoT-based personalized data construction process and ensured the factual accuracy of generated data through rigorous data cleaning. In the context of Douyin search ads, we added personalized title candidates for each ad campaign. Extensive offline experiments verified the effectiveness of our design, and online A/B testing demonstrated statistically significant gains.


\bibliographystyle{ACM-Reference-Format}
\bibliography{sample-base}
\clearpage

\appendix

\section{Ablation Study on Academic Dataset}
\label{sec:academic}
We constructed personalized title data based on the academic dataset Amazon Book Reviews~\cite{books} and conducted a series of experiments to demonstrate the effectiveness of HLLM-Creator. We use original title and description as item-side constraints. It should be noted that the constructed data was not subjected to hallucination filtering and was used solely for modeling validation of personalized generation, without practical significance.

\subsection{User Modeling Method}
\begin{figure}
  \centering
  \includegraphics[width=\linewidth]{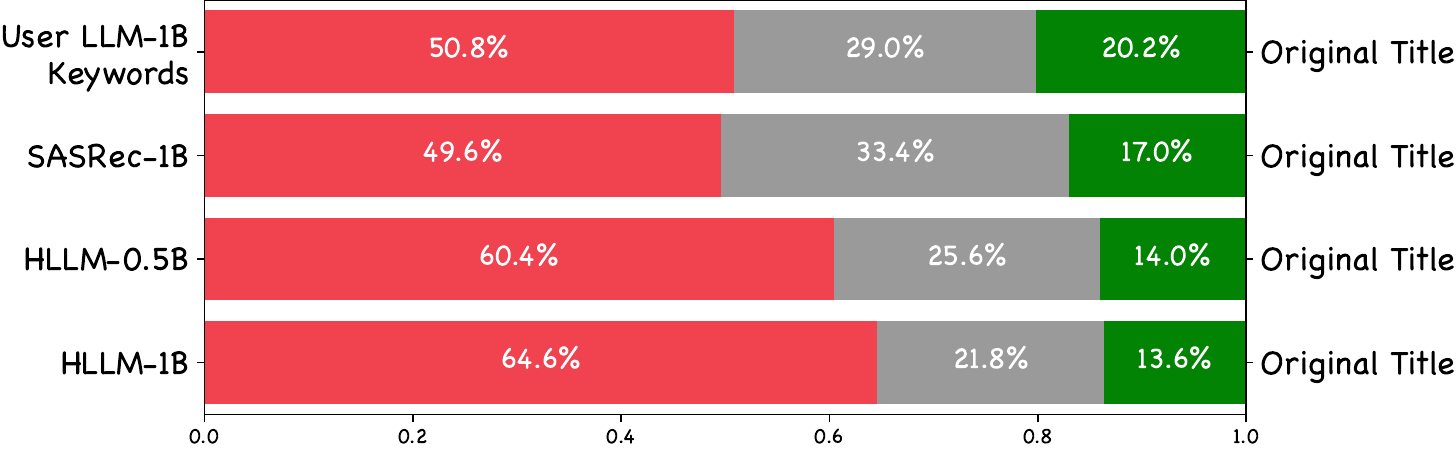}
  \caption{Comparison of HLLM-Creator with other user modeling methods on academic dataset.}
  \label{fig:comparison_academic}
\end{figure}

In the main text, we have verified the necessity of personalized user modeling. Here, we further validate this conclusion on the academic dataset, showing that introducing personalized modeling leads to significant improvements compared to using original titles. Additionally, we demonstrate that the quality of user modeling also has an important impact on personalized generation. As shown in Figure~\ref{fig:comparison_academic}, we first replace the Item LLM with keywords extracted from titles by LLM or with item IDs, labeled as User LLM-1B Keywords and SASRec-1B, respectively. Compared to the original titles, these approaches yield improvements of 30.6\% and 32.6\%, respectively, but are less advantageous than modeling user features with HLLM (51\%), denoted as HLLM-1B. Furthermore, the size of the HLLM model also affects the results: when the model size is reduced from 1B to 0.5B, the advantage decreases to 46.4\%.

\subsection{Auxiliary Loss}
We also validate the effectiveness of the auxiliary loss on the academic dataset, with the results presented in Table~\ref{tab:aux_loss_academic}, where the baseline is the original title. Consistent with the findings on the industrial dataset, HLLM-Creator's advantage increases from 51.0\% (without any auxiliary loss) to 59.4\%.

\begin{table}
  \caption{Ablation study of auxiliary losses on the academic dataset.} 
  \label{tab:aux_loss_academic}
  \begin{tabular}{cccrrrr}
    \toprule
    align & cls & recon  & Good & Same   & Bad & Advantage  \\ \midrule
                &             &             & 64.6\% & 21.8\%  & 13.6\%  & 51.0\% \\ 
    \usym{2713} &             &             & 67.4\% & 20.6\%  & 12.0\%  & 55.4\% \\ 
                & \usym{2713} &             & 67.6\% & 21.6\%  & 10.8\%  & 56.8\% \\ 
                &             & \usym{2713} & 65.2\% & 22.0\%  & 12.8\%  & 52.4\% \\ 
    \usym{2713} & \usym{2713} &             & 67.6\% & 22.8\%  & 9.6\%   & 58.0\% \\
    \usym{2713} & \usym{2713} & \usym{2713} & 70.6\% & 18.2\%  & 11.2\%  & \textbf{59.4\%} \\ \bottomrule
\end{tabular}
\end{table}

\subsection{Clustering}
\begin{figure}
  \centering
  \includegraphics[width=\linewidth]{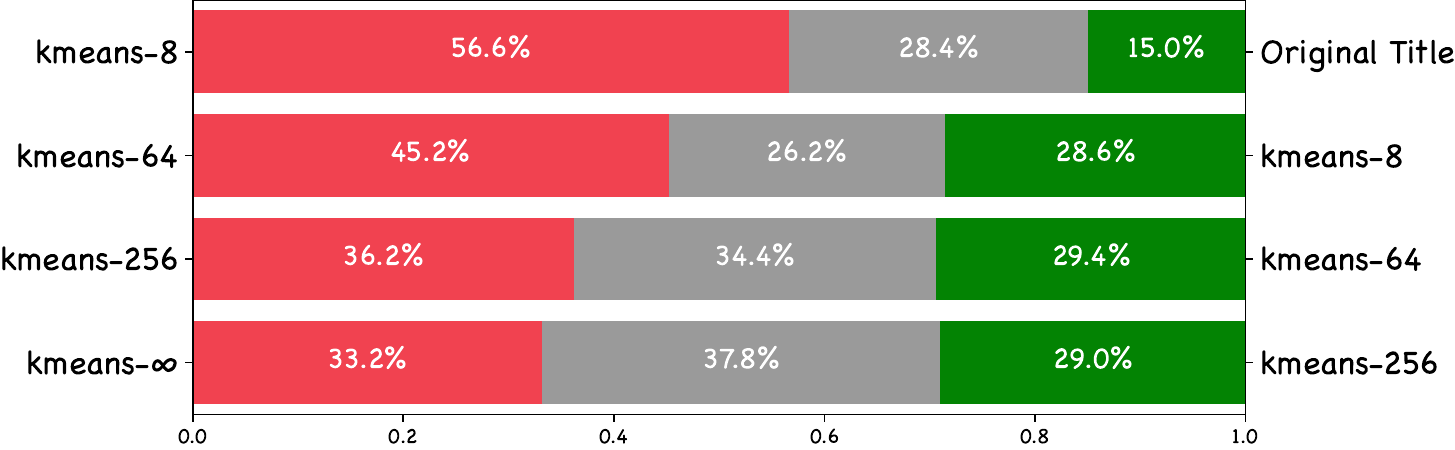}
  \caption{Ablation study of the impact of clustering on the academic dataset. $\infty$ indicates no clustering applied.}
  \label{fig:clustering_academic}
\end{figure}

Table~\ref{fig:clustering_academic} shows the impact of user clustering on the academic dataset. Similarly to the conclusions on the industrial dataset, user clustering is detrimental to the performance of generating personalized titles. As the number of clusters increases, the model performance gradually improves, with the best results achieved when no clustering is applied.

\section{Prompt Engineering}
\label{sec:prompt}

In Section~\ref{sec:dataset} and Section~\ref{sec:eval_metric} of the main text, we designed several prompts. Here, we provide the specific content of these prompts. For ease of reading, we have translated the Chinese into English.

\begin{figure*}
  \centering
  \includegraphics[width=0.94\linewidth]{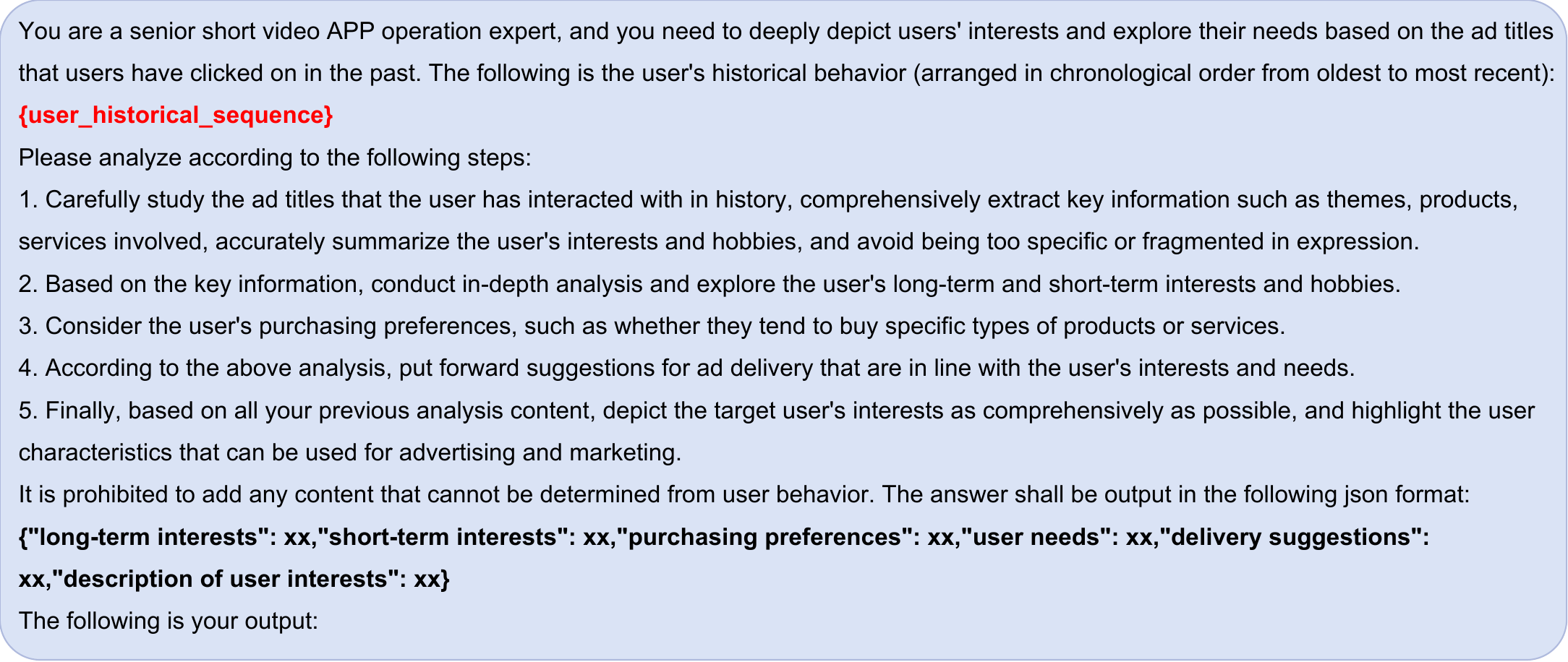}
  \caption{The prompt used for User Interest Profiling.}
  \label{fig:user_interest_extraction}
\end{figure*}

Figure~\ref{fig:user_interest_extraction} shows the prompt of User Interest Profiling, which covers user interests across multiple dimensions.  

\begin{figure*}
  \centering
  \includegraphics[width=0.94\linewidth]{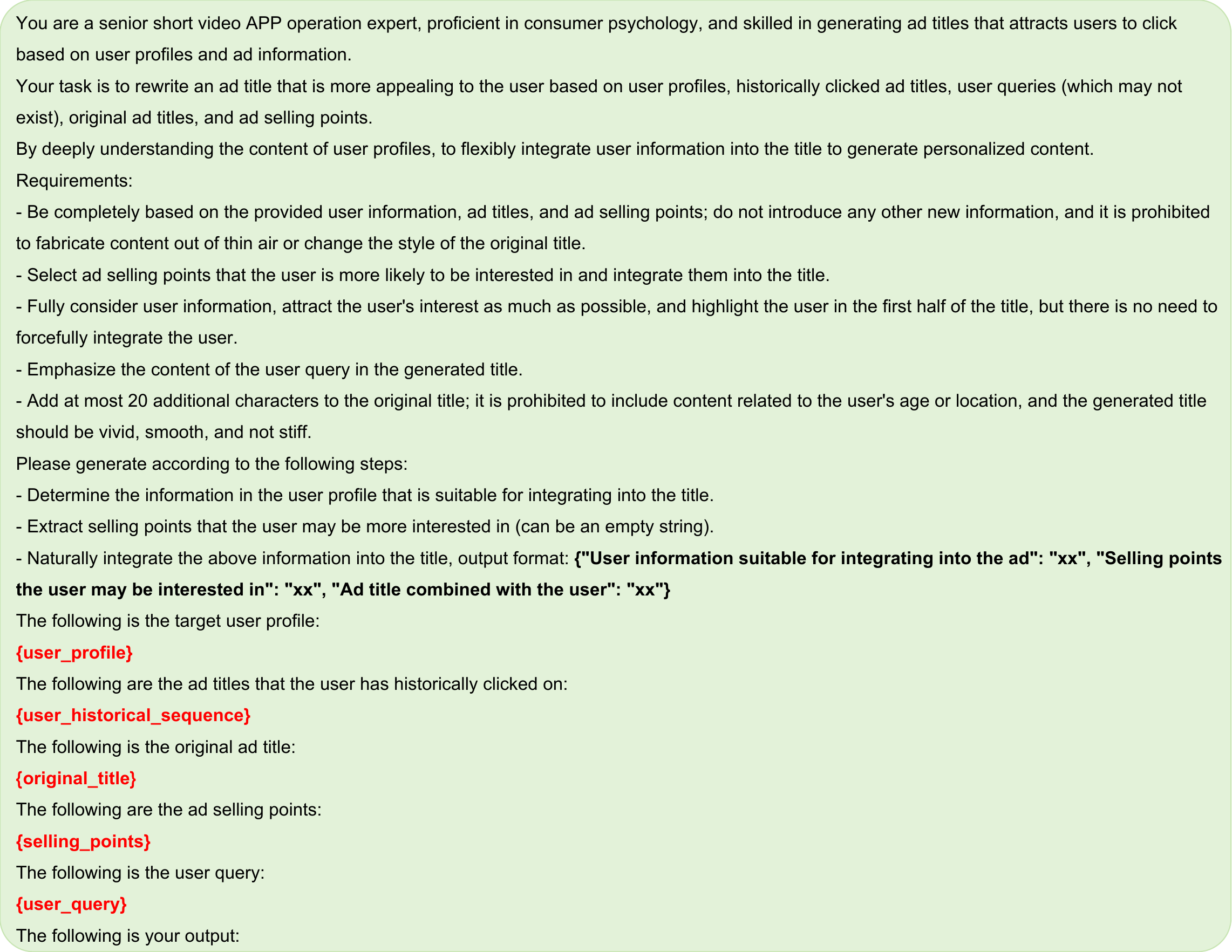}
  \caption{The prompt used for Interest-Driven Title Generation.}
  \label{fig:title_generation}
\end{figure*}

Figure~\ref{fig:title_generation} displays the prompt of Interest-Driven Title Generation.  

\begin{figure*}
  \centering
  \includegraphics[width=0.97\linewidth]{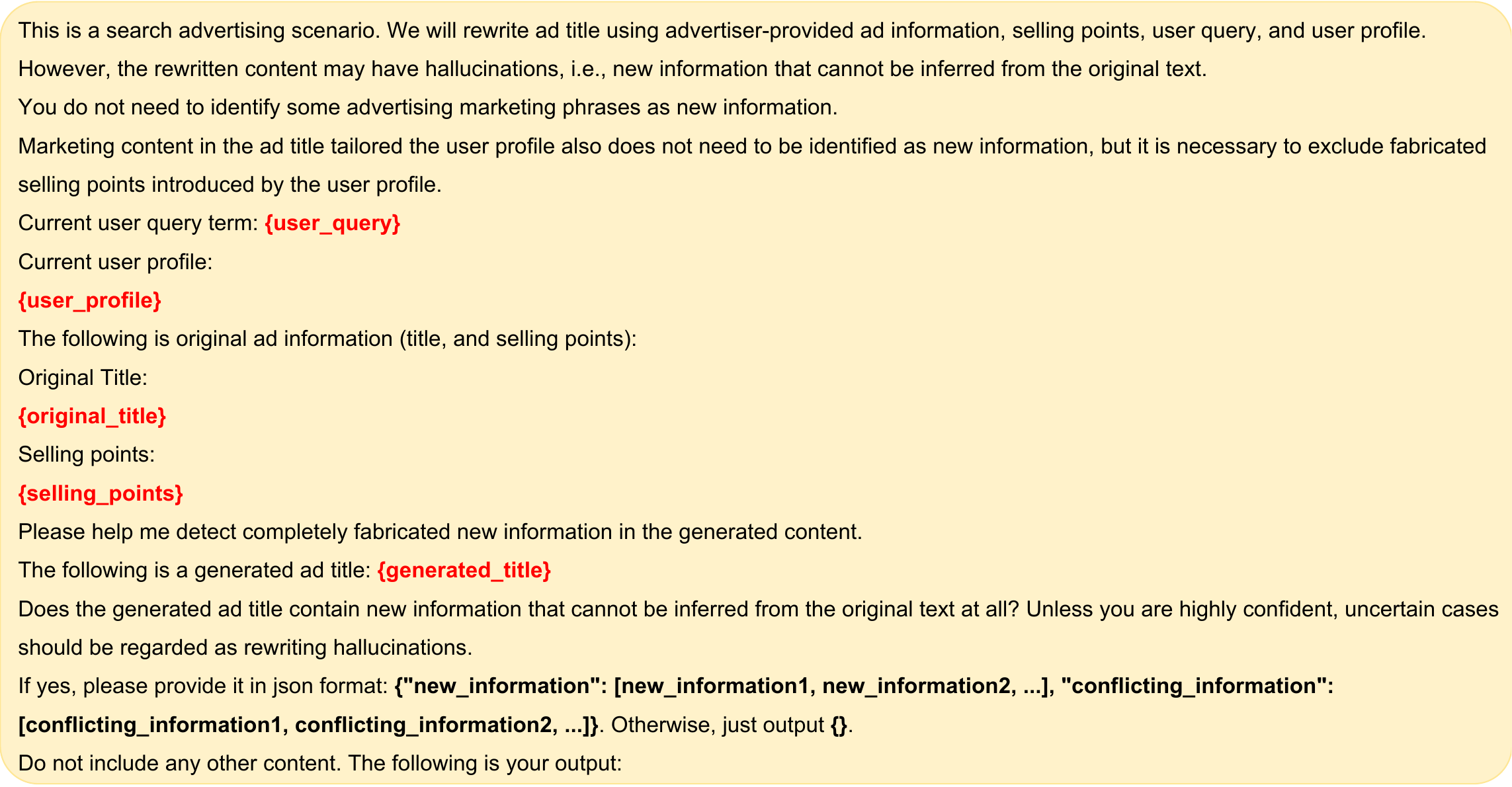}
  \caption{The prompt used for Hallucination-Free Title Filtering and hallucination evaluation. Only the output \textbf{"\{\}"} is considered to pass the evaluation.}
  \label{fig:data_clean}
\end{figure*}

Figure~\ref{fig:data_clean} presents the prompt of Hallucination-Free Title Filtering, which is also used in hallucination evaluation. By filtering out hallucination-containing data, we ensure the reliability of the personalized title generation model in industrial scenarios.  

\begin{figure*}
  \centering
  \includegraphics[width=0.97\linewidth]{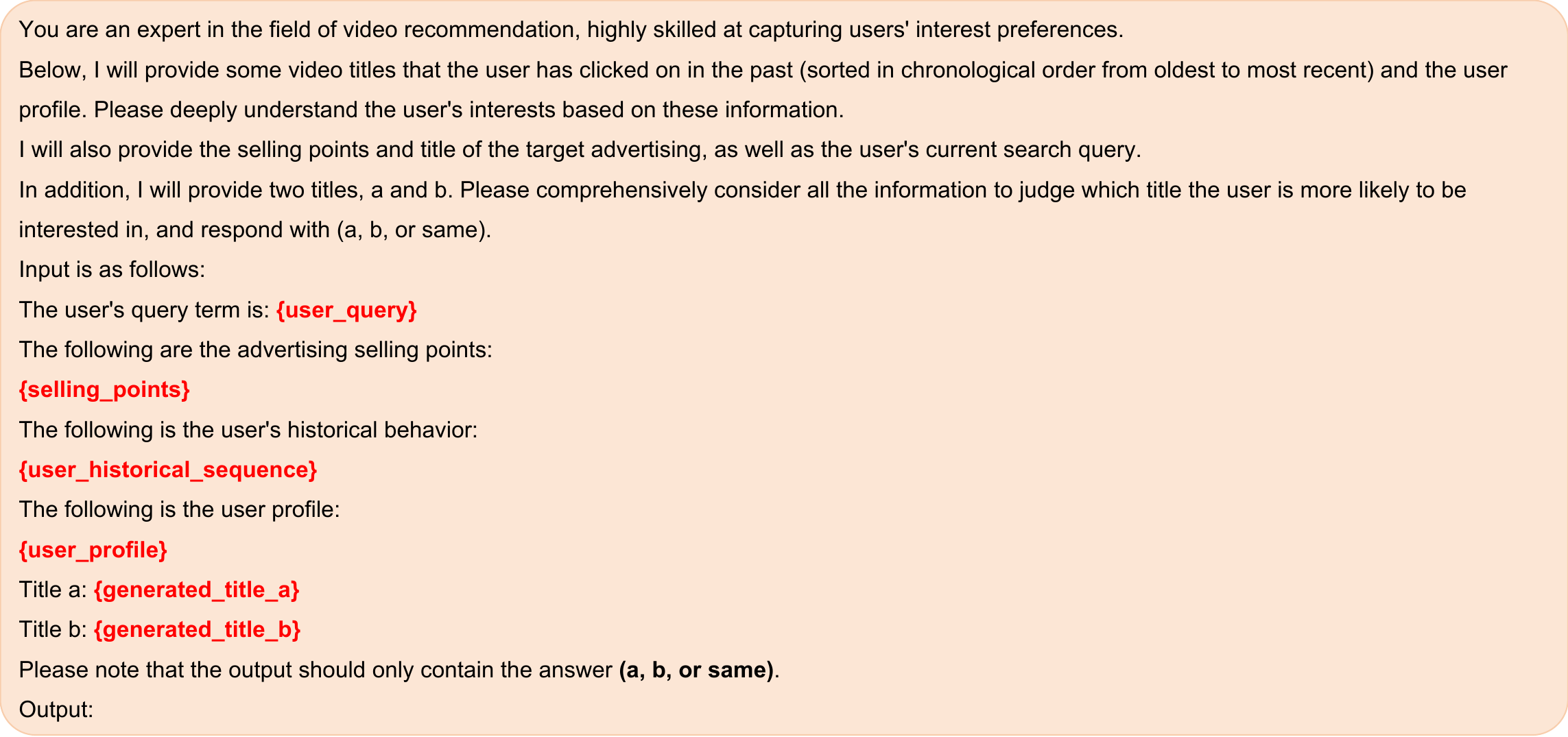}
  \caption{The prompt used for GSB performance evaluation.}
  \label{fig:eval_prompt}
\end{figure*}

Figure~\ref{fig:eval_prompt} shows prompts for offline model performance evaluation. By evaluating (model A, model B) and (model B, model A) pairs twice, we ensure the reliability of the evaluation results.

\section{Case Study} 
\label{sec:case_study}
Figure~\ref{fig:user_case1} and Figure~\ref{fig:user_case2} present the intermediate results of the CoT process (including user interests, user interests, and selling points suitable for inclusion in the title), ad information (original title and ad selling points), and the final generated personalized titles. For privacy protection, the cases shown here are generated by large language models and are not real data; they are for demonstration purposes only. However, the conclusions are generally consistent with those observed in real-world scenarios.

\definecolor{myblue}{RGB}{0,176,240}

\begin{figure*}
  \centering
  \includegraphics[width=\linewidth]{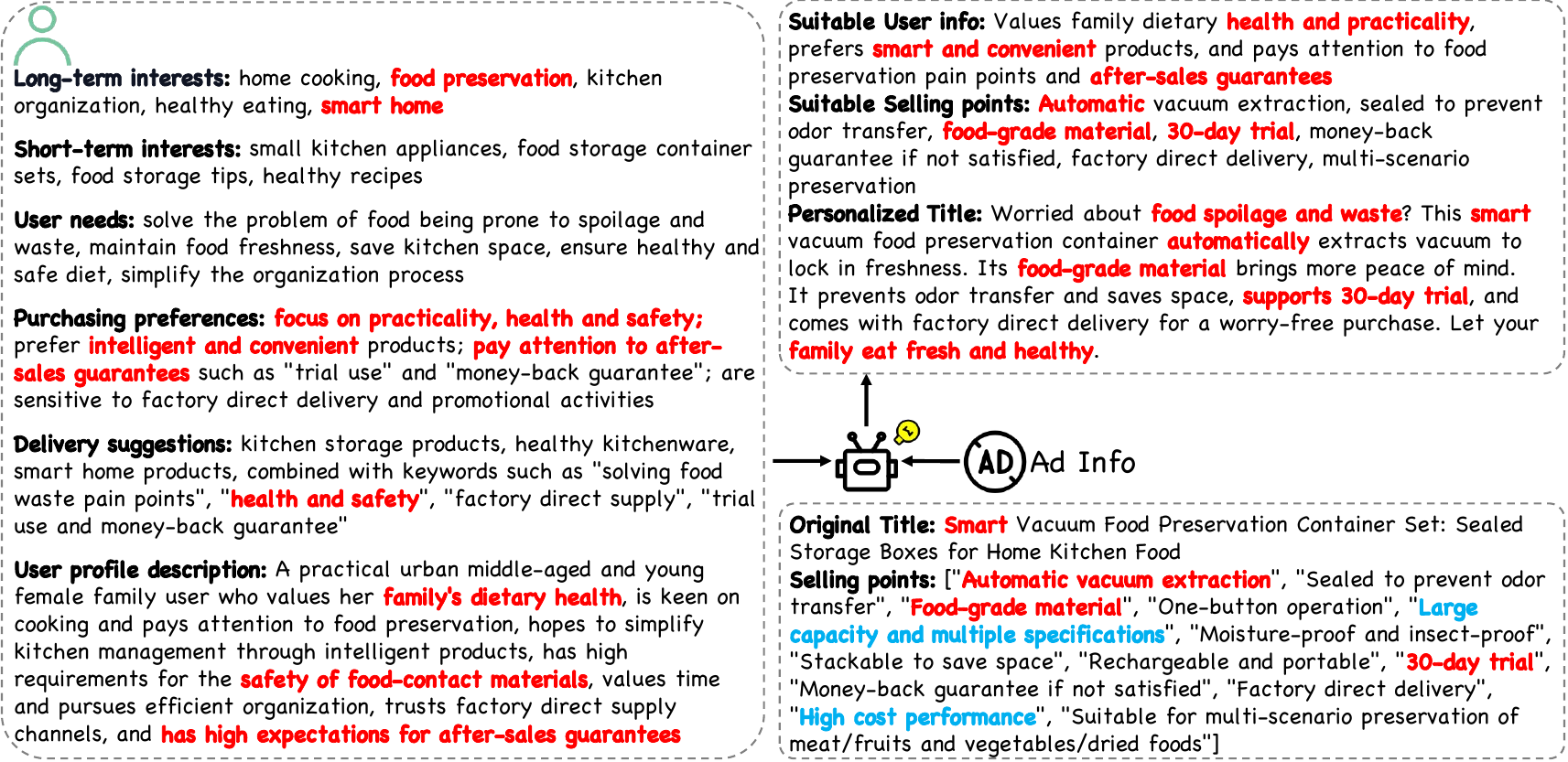}
  \caption{CoT-based personalized title, which extracts selling points more aligned with user interests such as "health", "smart", and "after-sales service". \textcolor{red}{Red text} indicates content that matches the user, while \textcolor{myblue}{Blue text} indicates content that does not match the user.}
  \label{fig:user_case1}
\end{figure*}

\begin{figure*}
  \centering
  \includegraphics[width=\linewidth]{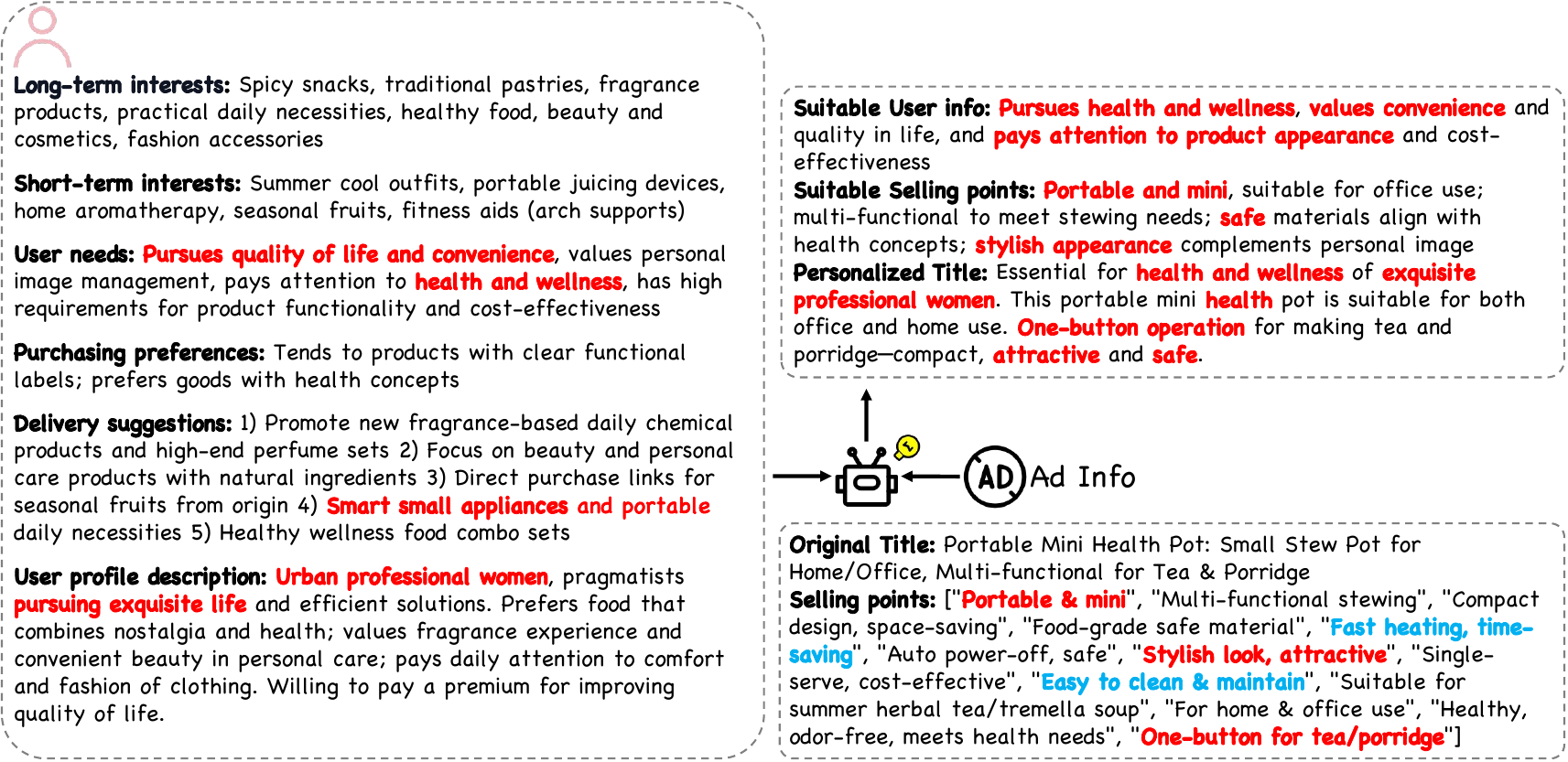}
  \caption{CoT-based personalized title, which explicitly emphasizes the user feature description "professional women". \textcolor{red}{Red text} indicates content that matches the user, while \textcolor{myblue}{Blue text} indicates content that does not match the user.}
  \label{fig:user_case2}
\end{figure*}

\end{document}